# Origin of Water in the Terrestrial Planets: Insights from Meteorite Data and Planet Formation Models

Andre Izidoro[1,2] and Laurette Piani[3]

**ABSTRACT**

Water condensed as ice beyond the water snowline, the location in the Sun's natal gaseous disk where temperatures were below 170 K. As the disk evolved and cooled, the snowline moved inwards. A low temperature in the terrestrial planet-forming region is unlikely to be the origin of water on the planets, and the distinct isotopic compositions of planetary objects formed in the inner and outer disk suggest limited early mixing of inner and outer Solar System materials. Water in our terrestrial planets has rather been derived from H-bearing materials indigenous to the inner disk and delivered by water-rich planetesimals formed beyond the snowline and scattered inwards during the growth, migration, and dynamical evolution of the giant planets.

## WATER CONTENT AND ISOTOPIC COMPOSITIONS OF PLANETARY MATERIALS

The origin and existence of water on Earth and the other terrestrial planets is a long-standing problem in planetary science. Liquid water covers more than 70% of the Earth's surface, yet this reservoir (hereafter "oceans") corresponds to only 0.025% of Earth's total mass. The main reservoir of Earth's water is most likely hidden in its interior (Peslier and De Sanctis 2022 this issue). Contrary to the Earth's surface, where hydrogen is mainly present as $H_2O$ molecules, "water" in Earth's interior corresponds to hydrogen incorporated in minerals, magmas, and fluids. During planetary evolution, hydrogen in these reservoirs can bond with surrounding oxygen to form water at the appropriate temperature and pressure conditions. Recent estimates based on mantle-derived rocks suggest that between one and seven Earth oceans' masses could be stored in the mantle. Earth's core could also contain a significant amount of hydrogen, although the exact quantity remains poorly constrained (FIG. 1A).

[1] Department of Earth, Environmental and Planetary Sciences
MS 126, Rice University
Houston, TX 77005, USA
E-mail: izidoro.costa@gmail.com

[2] Department of Physics and Astronomy
MS 550, Rice University
Houston, TX 77005, USA

[3] Centre de Recherches Pétrographiques et Géochimiques (CRPG)
Centre National de Recherche Scientifique (CNRS)
Université de Lorraine
Vandoeuvre-les-Nancy, F-54500, France
E-mail: laurette.piani@univ-lorraine.fr

Given the limited number of Martian meteorites, and the absence of Venusian and Mercurian meteorites, estimating the water content for the bulk Mars, Venus, and Mercury is rather difficult. The study of Martian meteorites indicates that the Martian mantle does contain water, although in smaller proportions than the Earth's mantle, and probably with a much more heterogeneous distribution. Liquid water was also once present on the Martian surface (Peslier and De Sanctis 2022 this issue). Venus could contain significant amounts of water in its mantle, in quantities comparable to that of Earth's mantle (McCubbin and Barnes 2019). It is likely that Venus shared a similar accretion history to Earth and was initially wet, although it probably remained in this state for only a short period of time (<1 Gy; Peslier and De Sanctis 2022 this issue). The presence of H, possibly in the form of pure water ice, has been detected at the surface of Mercury, and estimates based on the abundance of other volatile elements in its exosphere (He, Na, K) suggest that Mercury's mantle could contain as much water as the Martian mantle (McCubbin and Barnes 2019).

Not only planets carry water, but also minor planetary objects such as some asteroids. The terrestrial and giant planets are separated by the asteroid belt, a region populated by a large number of asteroids. Asteroids are thought to be leftovers from the birth of the Solar System and hold records of where and when planet formation started, and how the terrestrial planets obtained their water. The asteroid belt is the major source of meteorites found on Earth. Meteorite analysis and astronomical observations show that the asteroid belt is dichotomous. Rocky, water-poor asteroids associated with the parent bodies of non-carbonaceous chondrites (NCs) are more abundant in the inner belt (ordinary and enstatite chondrites groups in FIG. 1A; hereafter referred to as OCs and ECs, respectively) and dark, potentially volatile-rich asteroids thought to be made of similar materials as the carbonaceous chondrites (CCs) are more abundant in the outer belt and close to Jupiter's orbit (CC group in FIG. 1A; DeMeo and Carry 2014). This is consistent with the maximum amount of water measured in the different chondrite classes, which broadly increases with the putative orbital radius of their parent body (FIG. 1A). The NC and CC meteorites also have distinct isotopic compositions. There is a clear dichotomy in the distribution of neutron-rich isotopes, such as $^{54}Cr$ or $^{50}Ti$ (Kleine et al. 2020), which suggests that these two reservoirs of planetary materials (NC and CC groups) co-existed when the planets began to form, but remained physically separated during the first few million years of the Solar System's history. The close orbital proximity of asteroids associated with NC and CC meteorites in the asteroid belt today (DeMeo and Carry 2014) has thus to be explained by an additional process that broadly mixed these asteroid groups later (e.g., Izidoro et al. 2022).

Planets may either be born with water or have their water delivered by collisions with water-bearing planetary objects like asteroids. The H isotopic composition of water (expressed as the D/H ratio or in δ notation relative to Earth's standard mean ocean water) in planetary objects is a key discriminant to trace planets' main source(s) of water (Piani et al. 2021; FIG. 1B). The Sun's natal gas disk and Jupiter's gaseous envelope show very low D/H ratios of $\approx 21 \times 10^{-6}$ (or $\delta D \approx -800‰$). Earth's ocean D/H ratio is $\approx 156 \times 10^{-6}$ (or $\delta D \approx 0‰$), which is a factor of seven higher than that of Jupiter/Sun. Outer Solar System objects, such as the comet 67P/Churyumov-Gerasimenko (Bockelée-Morvan et al. 2015; FIG. 1B), generally show much higher D/H ratios in the sublimated water molecules, with values up to $530 \times 10^{-6}$ (or $\delta D \approx 2400‰$). Comets were historically considered to be the most likely source of Earth's water due to their high water content (about 50%) and water D/H ratio, sometimes similar to that of Earth's oceans (Bockelée-Morvan et al. 2015). However, if we consider molecules other than water, such as hydrogen cyanide HCN (FIG. 1B), which also are contained in cometary ices, comets would have an average D/H ratio systematically higher than Earth's oceans, making their contribution to the terrestrial planet water limited (Alexander et al. 2018). The CCs and OCs cannot be easily discriminated on the simple basis of their D/H ratios, as both chondrite classes cover large D/H ranges including that of Earth's oceans (FIG. 1B). D/H variations are also measured in minerals of achondrites, such as eucrites, angrites, or aubrites (Peslier and De Sanctis 2022 this issue). Both the NC and CC reservoirs thus seem to have sampled water with a wide D/H range that might result from different degrees of isotopic equilibration between an initially D-rich materials—partially preserved in cometary ices—and the D-poor solar gas (Piani et al. 2021).

In the next sections, we briefly review the current paradigm of planet formation and the implications for the nature of material prone to be incorporated into terrestrial planets during their formation.

## PLANETARY GROWTH PROCESSES: FROM DUST GRAINS TO PLANETS

The temperature of gaseous protoplanetary disks broadly decreases with heliocentric distance. Water ice can only be present in regions where the local temperature is lower than the water condensation temperature (~170 K), at the so-called water snowline (e.g., Morbidelli et al. 2016).

The first solid grains to condense in the disk had typical sizes of observed interstellar dust grains. They stick together forming millimeter- and centimeter-sized particles called pebbles. Pebbles drift via aerodynamic drag, inducing a flux of mass from the outermost to the innermost parts of the disk (e.g., Johansen and Lambrechts 2017). Pebble drift is a potential way to transfer water-rich solids formed beyond the snowline to the dry inner part of the Solar System (Morbidelli et al. 2016).

Numerical simulations suggest that planet formation "jumps" from millimeter-/centimeter-sized pebbles to ~100 km asteroidal-sized bodies (called planetesimals) due to particle–gas instabilities (Johansen and Lambrechts 2017). Finally, planetesimals grow to protoplanets either via accretion of other planetesimals ("planetesimal accretion") or of drifting pebbles ("pebble accretion"; Izidoro and Raymond 2018).

The interaction of a growing protoplanet with a mass of ~10 Earth masses (i.e., a giant planet core) with its surrounding gas induces a high gas pressure region just outside its orbit, called a "pressure bump" (Johansen and Lambrechts 2017), that tends to stop drifting pebbles. If the induced pressure bump is sufficiently strong, the forming planet stops growing via pebble accretion and is said to have reached the "pebble isolation mass". This pebble barrier may drastically affect the growth of any inner planets in the system by regulating the amount of drifting pebbles reaching the inner disk (e.g., Izidoro et al. 2021). The accretion history of terrestrial planets and their ability to accrete wet materials can thus be strongly affected by the formation of giant planets.

# DIFFERENT SCENARIOS FOR THE ORIGIN OF WATER ON TERRESTRIAL PLANETS

### *Oxidation of a Primordial H-rich Atmosphere*

As a planet grows, the heat provided by the radioactive decay of $^{26}$Al, accretionary impacts, and core formation tends to melt its surface and interior. Protoplanets with masses above ~0.5 Earth masses may be able to accrete gas from the disk and retain a gaseous atmosphere (Lammer et al. 2018). Water can be produced through oxidation of a hydrogen-rich atmosphere interacting with the planet's molten surface. This process may be an important source of water for terrestrial worlds around other stars (Kite and Schaefer 2021), but does not likely represent a major source of Earth's water. The D/H ratio of water produced via oxidation of a hydrogen-rich atmosphere should be similar to that of the Sun's natal disk. However, the D/H ratios of the Earth's oceans and Martian rocks are higher than solar (FIG. 1). Hydrogen hydrodynamic escape (Lammer et al. 2018), which was initially proposed as a potential mechanism to increase the D/H ratio of the primordial Earth from solar to the current value, does not seem to have been efficient on Earth (Sossi et al. 2020). This process might have been more important for Mars due to its lower gravity, and for Venus due to its formation location closer to the Sun (Peslier and De Sanctis 2022 this issue).

### *Endogenous H-bearing Materials in the Inner Disk*

Even if the disk temperatures did not drop below 170 K in the terrestrial region, hydrogen-bearing solids other than water ice could have contributed to the final water budget of terrestrial planets. This is consistent with the observation of H-bearing phases—organics and anhydrous silicates—in ECs and OCs thought to have formed in the inner disk (Piani et al. 2020; Jin and Bose 2021).

The chemical bonding of water vapor molecules at the surface of solid dust grains allows water to sustain temperatures ≥ 500 K. This adsorbed water was proposed to have provided substantial amounts of water to the terrestrial planets (up to three oceans to the Earth; Drake 2005). Similarly, $H^+$ implantation in the interstellar medium, or interplanetary silicate grains by ionized nebular gas or solar wind, may have produced trace amount of hydroxyl ($OH^-$) or water in the dust, eventually accreting on terrestrial planets (e.g., Jin and Bose 2021). The low D/H ratios of implanted nebular gas or solar wind may not be consistent with that of primitive (undifferentiated) meteorites or terrestrial planets. However, the extent of the isotopic fractionation associated with the ionization/implantation mechanisms is not yet fully understood (Jin and Bose 2021).

### *Snowline Sweeping*

Young gaseous protoplanetary disks are initially very hot and massive. As the disks age, they lose mass and cool. Silicates and water condense at temperatures of about 1400 and 170 K, respectively. In realistic disk models, the snowline is initially around ~10 to ~20 astronomical units (AU) and the silicate line is initially between ~0.1 and ~1 AU (e.g., Morbidelli et al. 2016; Izidoro et al. 2022). Planetesimals and protoplanets that formed entirely from material accreted beyond the snowline are probably born with large water mass fractions. As water ice condenses, the gas beyond the snowline becomes dry. The gas in the disk accretes onto the star moving inwards faster than the snowline. Gas moving from beyond the snowline has no more water vapor to condense (Morbidelli et al. 2016), thus, the delivery of water to the inner regions of the disk requires the drift of pebbles originating from beyond the snowline (Morbidelli et al. 2016). If the water snowline eventually reaches the terrestrial planet region, icy pebbles may either deliver water to pre-existing dry planetesimals and protoplanets or promote the formation of new water-rich planetesimals (Lichtenberg et al. 2021). Water delivery via sweeping of the water snowline may result in Earth-like water contents if pebble accretion is not efficient or if ice pebbles mostly evaporate in the planet's atmosphere (Johansen et al. 2021), getting recycled in the disk rather than being accreted. Although snowline sweeping may have delivered water to rocky planets around other stars (FIG. 2A), it is unlikely to have played a role in the Solar System. The isotopic dichotomy between NC and CC meteorites suggests that a pressure bump at the snowline or Jupiter's early formation efficiently prevented water-ice pebbles from the outer Solar System from drifting into the inner Solar System (Brasser and Mojzsis 2020; Kleine et al. 2020; Burkhardt et al. 2021; Izidoro et al. 2021). Without an efficient pebble barrier, aerodynamic drift would have largely mixed inner and outer Solar System materials. This mixing should have resulted in the formation of planetary objects with isotopic compositions intermediate between NC and CC, but this view is inconsistent with meteorite data. Therefore, even if the snowline eventually reached the terrestrial region, water-rich ice pebbles probably did not (Morbidelli et al. 2016; Izidoro et al. 2022). It thus seems very unlikely that snowline sweeping delivered water to the terrestrial planets.

*Inward Migration of Water-rich Protoplanets*

Moon- to Mars-mass protoplanets (and less massive objects such as planetesimals) are unlikely to have experienced large-scale, gas-driven, radial migration. Earth-mass planets, on the other hand, are massive enough to strongly interact with the gas and migrate typically inwards. If water-rich planets form beyond the snowline, they may eventually migrate inwards, parking at or crossing the terrestrial region (Izidoro et al. 2021). Jupiter's early formed core or a pressure bump at the snowline is believed to have stalled the growth of the terrestrial planets via pebble accretion before they could become massive enough to migrate significantly inwards, potentially to regions of the disk much closer to the star (Izidoro et al. 2022).

Planetary gas-driven migration has been proposed to explain the origin of the so-called hot super-Earths (e.g., Izidoro et al. 2021). Super-Earths are planets with masses between those of Earth and Neptune. Statistical analyses suggest that 30%–50% of Sun-like stars host hot super-Earths with orbital periods shorter than 100 days. The migration hypothesis for the origin of super-Earths is consistent with a broad range of super-Earth compositions, from dry planets to water worlds, depending on whether planetesimals/protoplanets form more efficiently inside or outside of the water snowline (Izidoro et al. 2021).

Some of the best characterized super-Earth systems may have planets holding water, although estimates come with very large error bars (e.g., Raymond et al. 2021). The TRAPPIST-1 is one of these iconic systems in which adjacent planets show rhythmic orbital motions and form a so-called "resonant chain". In simple terms, resonance occurs if two planets consistently complete an integer number of orbits around the star at the same time. This orbital rhythm is not coincidental, but is the result of gas-driven migration (e.g., Izidoro et al. 2021). The current dynamical configuration of the TRAPPIST-1 system probably represents its dynamical state at the end of the gas disk phase, several billion years ago. If any planet in the TRAPPIST-1 system contains a large water budget (i.e., ~10 wt.%), this implies that it must have been incorporated during its formation in the gaseous disk. It is unlikely that a large water reservoir could be delivered to this system via impacts with objects from beyond the snowline (e.g., asteroids, protoplanets, and/or comets; see next section), because the very sensitive dynamical configuration of the resonance chain would have been definitely lost (Raymond et al. 2021). The TRAPPIST-1 system and other resonant chains of hot super-Earths, such as TOI-178 and KEPLER-223, likely did not experience substantial water delivery after disk dissipation. Endogenous sources may have contributed at some level to their water budget. However, if these planets formed inside the snowline, by analogy with inner Solar System materials, one would naïvely expect them to have a relatively low water content, potentially lower than 1 wt.%.

*External Delivery*

Protoplanets and planetesimals on almost circular orbits can only collide with nearby planetesimals because of limited radial excursion when they orbit the star. Planetary objects on eccentric orbits—like those of comets—spend time both in the innermost and outermost regions of their planetary systems. If water-rich protoplanets or planetesimals formed beyond the snowline are dynamically excited to high eccentricity orbits, they may eventually cross the terrestrial region and deliver water to terrestrial planets via impacts (e.g., Meech and Raymond 2020).

Our current understanding of planet formation and Solar System dynamical evolution allows multiple episodes of water delivery to the terrestrial planets (see also Meech and Raymond 2020), spanning events taking place during the disk phase and potentially post-disk dissipation (Fig. 3).

A giant planet core that has just reached pebble isolation mass (i.e., >10–15 Earth masses; see above) accretes gas from the disk slowly. When the mass of the core's envelope is comparable to the mass of the core, "runaway" gas accretion begins and the planet mass doubles over short timescales. As a consequence, this fast-growing planet destabilizes nearby planetesimals and scatters them onto eccentric orbits (FIG. 3A–C; Izidoro and Raymond 2018). As Jupiter and Saturn started runaway accretion, a fraction of the planetesimals reaching the inner Solar System were implanted into the asteroid belt via gas-drag assistance. Another fraction was scattered even closer in, inside the water delivery zone (FIG. 3), and collided with the terrestrial planet building blocks delivering water.

The low-mass atmospheres of Uranus and Neptune suggest that they either formed later than Jupiter's and Saturn's cores or more slowly, and avoided becoming gas giants (FIG. 3A–C). As these putative protoplanets beyond the orbit of Saturn grew massive enough, they started to migrate inwards before being stopped by Jupiter and Saturn (FIG. 3D–E). The process of migration and accretion of Neptune-mass ice planets (FIG. 3F) sculpted the disk of planetesimals beyond the gas giant planets and scattered planetesimals from the distant regions of the disk into the inner Solar System (Ribeiro de Sousa et al. 2020). This provided another episode of external delivery of water to the terrestrial planets and the implantation of planetesimals into the belt region.

The most successful models of Solar System evolution—those that explain fairly well the current orbits of the four giant planets and orbital architecture of the objects in the very outer Solar System—suggest that the Solar System was born with three ice giants, but that one of them was ejected during the giant planet dynamical instability (FIG. 3G; Nesvorný 2018). The relatively violent dynamical evolution of the planets, again, scattered planetesimals from the most distant regions of the Solar System into the terrestrial region (Nesvorný 2018). Some of this material potentially had a water content and composition similar to those of comets (Nesvorný 2018). The exact timing of when the instability happened is not constrained, but several dynamical arguments suggest that it probably happened during the first 100 My of the Solar System history (Ribeiro de Sousa et al. 2020; Nesvorný 2018), when the terrestrial planets were still forming (FIG. 3).

### *A Hybrid Scenario for the Origins of Earth's Water*

The Earth's water isotopic composition is not homogeneous, with the most pristine mantle signatures being depleted in deuterium compared to the Earth's oceans (Peslier and de Sanctis 2022 this issue). Enstatite chondrites represent a good match to the D/H ratio of the Earth's primitive mantle (FIGS. 1 and 4) and could thus indicate the contribution of an endogenous source to Earth's mantle water. Enstatite chondrites also have nitrogen isotopic compositions that fit those inferred for the primitive Earth's mantle (FIG. 4; Piani et al. 2020), indicating that not only water but also other volatile elements might be derived from this endogenous source.

Isotopic analysis suggests that the Earth mostly formed from NC material possibly derived from the same reservoir as ECs (Kleine et al. 2020), which accreted inside the snowline (e.g., Izidoro et al. 2022; Morbidelli et al. 2022).

It is not trivial, however, to estimate the amount of water that this endogenous material could have provided to Earth. If one assumes that the Earth is made entirely of ECs, they could have accounted for 3 to 23 oceans of water (Piani et al. 2020). On the other hand, it is not clear whether Earth is made from the same EC-like reservoir that we have in our meteorite samples, or from an early population that experienced more extensive internal heating—due to the decay of radioactive nuclides such as $^{26}$Al—and differentiation, resulting in significative water loss (Peslier and de Sanctis 2022 this issue). If this latter hypothesis is correct, using the water content of ECs to estimate the contribution of endogenous material to the Earth's water may not be ideal because one may overestimate their contribution. Recent estimates of the water content in achondrites indicate a much lower water mass fraction than that estimated for pristine ECs (e.g., ≤0.007 wt.% of water in the Vestan mantle; Stephant et al. 2021 and references therein), which may reflect the effect of post-accretion processes dehydrating planetesimals. If one takes, for simplicity, this estimate as representative of the inner Solar System material that formed the Earth, one would expect a water contribution of roughly ≤0.3 oceans. Stronger constraints on the water content of both Earth's interior and differentiated meteorites would be more than helpful to further investigate this issue (McCubbin and Barnes 2019).

In any case, the Earth's water is unlikely to have been derived uniquely from NC material. Although the mantle isotopic composition strongly suggests that at least some water came from an endogenous source, the surface reservoirs of the Earth (oceans and atmosphere) are enriched in heavy isotopes compared to the mantle and ECs (FIG. 4). These isotopic compositions could be accounted for by considering the addition of some D- and $^{15}$N-rich CC materials to Earth during its accretion history. It is very unlikely that this contribution came from CC pebbles (e.g., Burkhardt et al. 2021), but rather from CC planetesimals that formed beyond the orbit of Jupiter and scattered into the inner Solar System (FIG. 3; Izidoro et al. 2022). A recent model invoking constraints from multiple isotopic systems suggests that Earth and Mars could have accreted up to ~4% of their current mass from a CC-like isotopic reservoir (Burkhardt et al. 2021), in agreement with previous estimates for Earth (e.g., Marty et al. 2016; Alexander et al. 2018). If one assumes that, on average, CC-like planetesimals carry a 5% water mass fraction (FIG. 1), it implies that the Earth and Mars received ~8.5 and ~0.85 Earth oceans from this reservoir, respectively, which is consistent with dynamical models (FIG. 3; Izidoro and Raymond 2018). Cometary materials could also have contributed to the late accretion of volatile-rich material, but mass-balance calculations including nitrogen and noble gases indicate that a cometary contribution to Earth's water was probably limited to a few percent (Marty et al. 2016), which is also in agreement with dynamical models of Solar System evolution (FIG. 3). Terrestrial planets in the Solar System most likely had a hybrid contribution of water both from endogenous and exogenous sources.

## CONCLUSION

Our understanding of the origins of water on terrestrial planets is far from complete, but significant progress has been made in the last decade or so. Terrestrial planets in the Solar System probably carry water derived from both inner and outer Solar System materials. The Solar System's giant planets' growth, migration, and dynamical evolution played a key role in delivering water to the inner Solar System via inward scattering of water-bearing planetesimals from the outer Solar System. Jupiter and Saturn probably also controlled water delivery to the inner Solar System by preventing water delivery to the terrestrial region via snowline sweeping (accretion of ice pebbles) and blocking the inward migration of water-rich worlds. The very contribution of different viable sources and timing of deliveries remain to be precisely determined. Further characterization of the overall composition of Earth-mass exoplanets around other stars, in particular, in well-characterized resonant chains, such as TRAPPIST-1 and TOI-178, is key to further improving planet formation theories and our understanding of the Solar System formation, including the origin of Earth's water.

## ACKNOWLEDGMENTS

We are very grateful to the reviewers, Alessandro Morbidelli and David Bekaert, for their thoughtful comments and suggestions, which tremendously helped us improve the quality of our paper. We are also thankful to Sean N. Raymond for carefully reading our paper and his helpful comments. A. Izidoro acknowledges The Welch Foundation grant no. C-2035-20200401 and NASA grant 80NSSC18K0828 for financial support during the preparation and submission of the work. A. Izidoro also thanks the CAPES-PrInt Program, process number 88887.310463/2018-00, International Cooperation Project number 3266, and CNPq (313998/2018-3). L. Piani acknowledges the French Research National Agency (grant ANR-19-CE31-0027-01 to L.P.). This is CRPG contribution 2814.

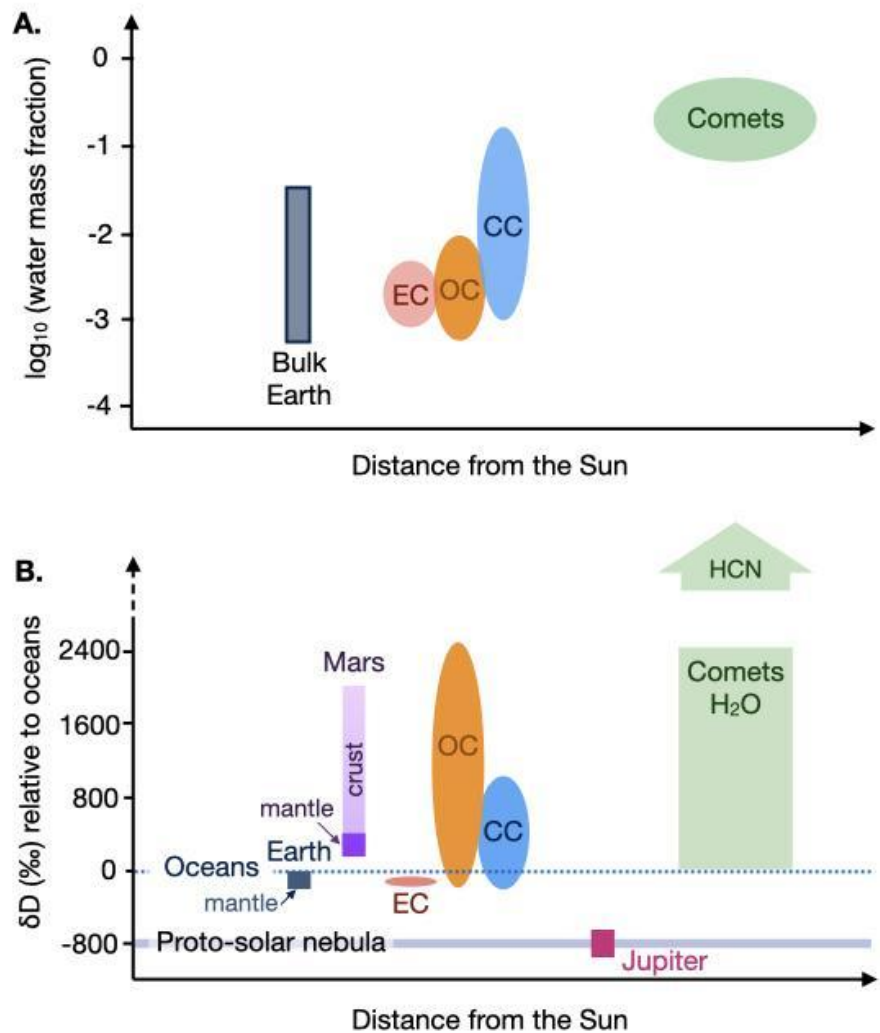

**FIGURE 1.** (**A**) Water content and (**B**) hydrogen isotopic composition of different planetary materials with putative heliocentric distances of their parent bodies. Water concentrations are reported in $H_2O$ mass fraction regardless of the nature of bonds between H and the other elements such as O, C, or N. Panel (**A**) shows a general increase in water content with increasing heliocentric distance. The total water content of Earth is estimated to be $3900^{+32700}_{-3300}$ ppm weight $H_2O$ (grey rectangle) (Peslier and De Sanctis 2022 this issue). For chondrites, the vertical height of the ellipsoids represents the range of water abundances measured for each chondrite class (Piani et al. 2021). Panel (**B**) illustrates the large variability of D/H ratios in different bulk chondrites (Vacher and Fujiya 2022 this issue) and comets (Bockelée-Morvan et al. 2015) in comparison to the D/H composition of terrestrial and Martian rocks (Peslier and De Sanctis 2022).

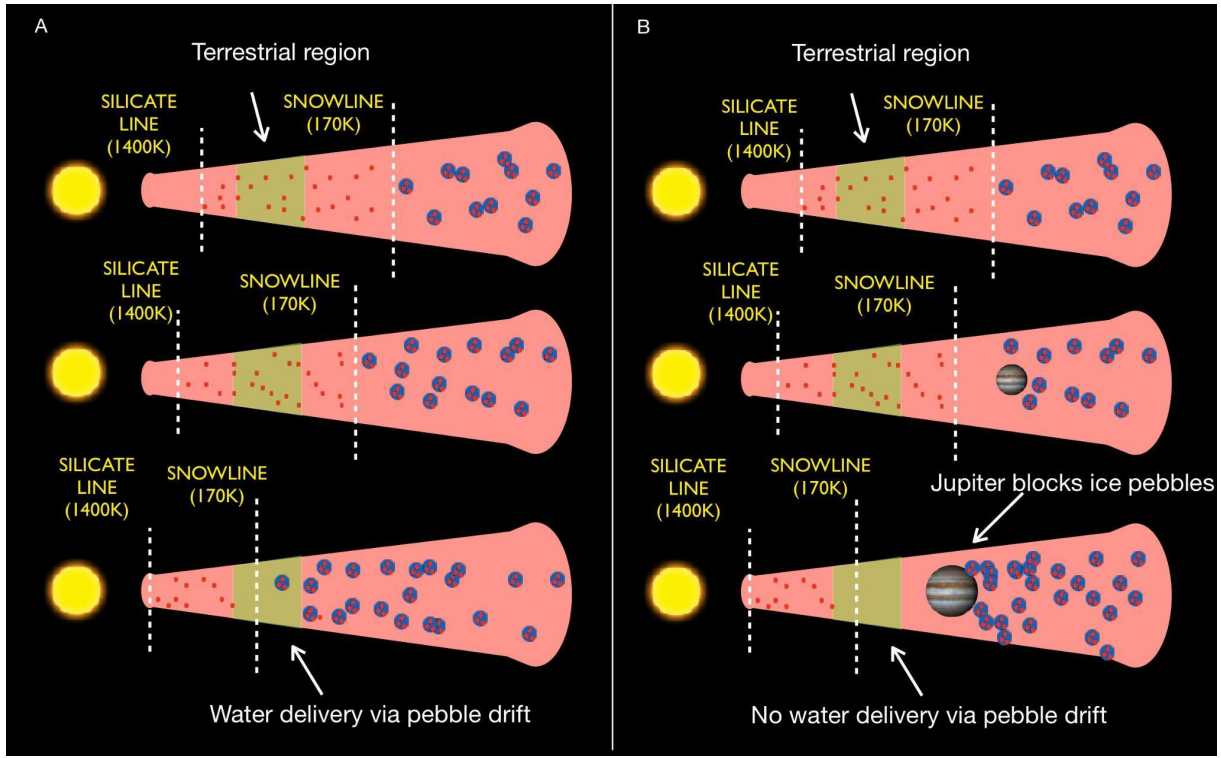

**FIGURE 2.** Evolution of protoplanetary disks around young stars without (**A**) and with (**B**) gas giant planets. Pebbles inside and outside the snowline are water-poor (small red dots) and water-rich (large blue-red dots), respectively. (**A**) As the disk evolves, the silicate line and snowline move inwards. Ice pebbles beyond the snowline drift and eventually reach the terrestrial region, either delivering water to pre-existing planetesimals/protoplanets or inducing the formation of new water-rich planetesimals. (**B**) The giant planet induces a pressure bump beyond its orbit, which prevents water-rich pebbles from drifting into the inner system. This evolution prevents water delivery via snowline sweeping. A disk with no giant planet but a pressure bump at the snowline line (pebble trap) would lead to an equivalent outcome (Izidoro et al. 2022).

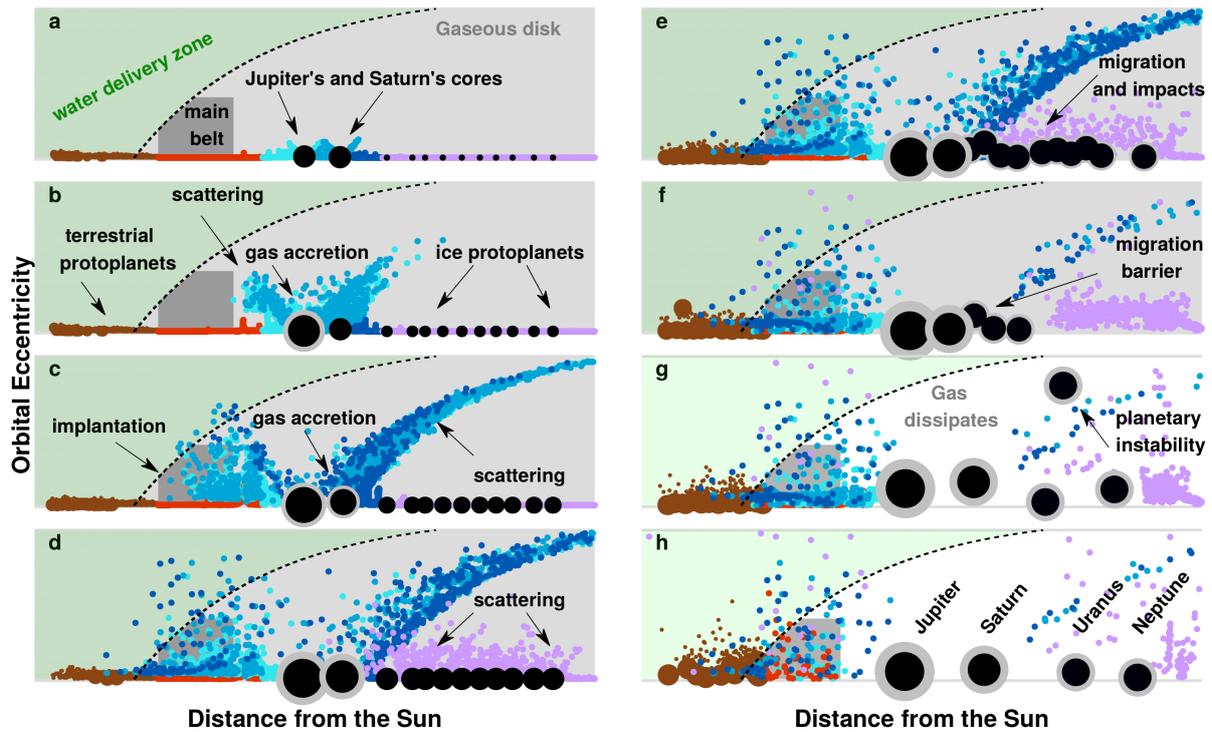

**FIGURE 3.** Dynamical evolution of the Solar System and different episodes of water delivery to the terrestrial planets. (**A**) Jupiter's and Saturn's cores grow via pebble accretion. (**B**) Jupiter's core starts to accrete its gaseous envelope from the disk and scatter nearby planetesimals towards the inner and outer Solar System. (**C**) Saturn's core starts to accrete its gaseous envelope and scatter planetesimals. (**D**) A fraction of the planetesimals scattered by Jupiter and Saturn is implanted into the asteroid belt, while another fraction reaches the terrestrial planets' water delivery zone (green region), delivering water to the growing terrestrial planets. (**E**) Ice protoplanets beyond Saturn grow via pebble and planetesimal accretion, migrate inwards due to planet-disk gravitational interactions scattering planetesimals, and eventually collide with other ice protoplanets. (**F**) Collisions among ice protoplanets lead to the formation of three ice giant planets beyond Saturn. (**G**) The sun's natal disk dissipates. In the absence of dissipative forces from the disk, the giant planets' orbits become dynamically unstable, leading to the ejection of one of the ice giants from the Solar System. The dynamical instability sculpts the planetesimal disk beyond the orbit of Neptune and scatters a fraction of these objects (magenta planetesimals) towards the inner Solar System. (**H**) The Solar System giant planets reach their current orbits and the terrestrial planets continue to grow. Terrestrial planets complete formation roughly 30 to 150 My after the beginning of the Solar System formation (not illustrated).

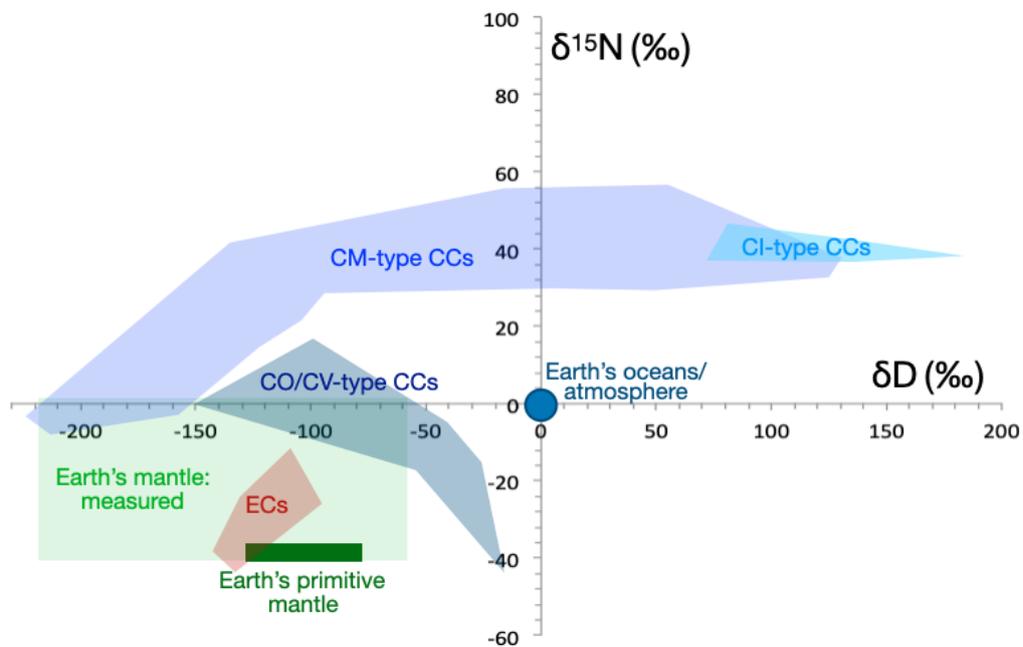

**FIGURE 4.** Hydrogen and nitrogen isotopic composition ranges of enstatite chondrites (ECs) and of different types of water-rich carbonaceous chondrites (CCs), namely the CMs (Mighei-type), CIs (Ivuna-type), CVs (Vigarano-type), and COs (Ornan-type), compared with the isotopic composition of the Earth's mantle and surface reservoirs. ADAPTED FROM PIANI ET AL. (2020).